\documentclass[]{spie}  %>>> use for US letter paper
\usepackage{amsmath,amsfonts,amssymb}
\usepackage{graphicx}
\usepackage{epstopdf}
\usepackage[colorlinks=true, allcolors=blue]{hyperref}

\title{Optimizing deep-space optical communication\\under power constraints}

\author[a]{Marcin Jarzyna}
\author[b]{Wojciech Zwoli\'{n}ski}
\author[b]{Micha{\l} Jachura}
\author[a,b]{Konrad Banaszek}
\affil[a]{Centre of New Technologies, University of Warsaw, Banacha 2c, 02-097 Warszawa, Poland}
\affil[b]{Faculty of Physics, University of Warsaw, Pasteura 5, 02-093 Warszawa, Poland}

\authorinfo{Further author information: (Send correspondence to K.B.)\\M.J.: E-mail: m.jarzyna@cent.uw.edu.pl\\  K.B.: E-mail: k.banaszek@cent.uw.edu.pl}

\begin{document}
\maketitle

\begin{abstract}
We investigate theoretically the efficiency of deep-space optical communication in the presence of background noise. With decreasing average signal power spectral density, a scaling gap opens up between optimized simple-decoded pulse position modulation and generalized on-off keying with direct detection. The scaling of the latter follows the quantum mechanical capacity of an optical channel with additive Gaussian noise. Efficient communication is found to require a highly imbalanced distribution of instantaneous signal power. This condition can be alleviated through the use of structured receivers which exploit optical interference over multiple time bins to concentrate the signal power before the detection stage.
\end{abstract}

% Include a list of keywords after the abstract

\section{INTRODUCTION}
\label{Sec:Introduction}

Using optical- rather than radio-frequency range for satellite communication prospectively offers higher bandwidths for data transfer, reduces diffractive losses in signal transmission, and additionally alleviates certain regulatory issues. However, in the deep-space regime the diminishing power of the received signal requires different signal modulation methods  compared e.g.\ to conventional fiber-optic communication.
%, where data are typically keyed using a constellation of complex field amplitudes.
A widely used format in deep-space communication is pulse position modulation (PPM), which encodes information in the temporal position of a light pulse within a frame of otherwise empty time bins.\cite{Hemmati,HemmatiBiswasProcIEEE2011} The standard technique to detect the PPM signal relies on time-resolved photon counting. Registering a photocount identifies the position of a pulse, while no photocount event within the frame implies erasure of the prepared PPM symbol. Such erasures can be dealt with efficiently using suitable forward error correcting codes.

The purpose of this paper is to identify the attainable efficiency of the PPM format under the constraint of a specified average signal power. The analysis includes counts generated by background noise assuming an arbitrary relation between the signal and the noise power, which goes beyond previously obtained results for noiseless or moderate-noise scenarios.\cite{WasedaSasakiJOCN2011,KochmanWangTIT2014,JarzynaKuszajOPEX2015,JarzynaBanaszekICSOS2017}
Optimization is carried out with respect to the length of the PPM frame.
Two statistical models for background counts motivated by different physical photodetection setups are considered. The Poissonian model is applicable to standard %time-integrated
photodetection, while the Gaussian model would be relevant to more advanced mode-selective detection. In the latter scheme the temporal mode carrying the modulated signal is separated prior to direct detection using e.g.\ nonlinear optical frequency conversion.\cite{EcksteinBrechtOpEx2011,BrechtEcksteinNJP2011,ReddyRaymerOpEx2013,BrechtReddyPRX2015,ReddyRaymerOpEx2017}
The efficiency limits are quantified with the help of Shannon mutual information which sets the theoretical upper bound on the attainable transmission rate for a given modulation/detection technique.\cite{CoverThomas} Shannon mutual information can be in principle attained with the help of forward error correction, although the efficiency of practical codes usually remains below that value.

In the presence of background noise the timing of a photocount does not necessarily match the position of a pulse. Furthermore, photocounts may occur in multiple time bins within a single PPM frame. The simple decoding strategy extracts information only from events when one photocount has been registered over the PPM frame, treating multiple photocounts as erasures. In this scenario mutual information exhibits unfavorable, quadratic scaling with the signal power for a fixed noise power \cite{Moision2014}. The PPM format can be viewed as a keying scheme using a binary constellation of elementary single-bin symbols ``on'' (a pulse) and ``off'' (an empty time bin) with a constraint that the ``on'' symbol is used exactly once in each predefined frame of time bins. In order to investigate the full potential of binary constellations for deep-space communication, generalized on-off keying (OOK) is considered in which the ``on'' and ``off'' symbols can be chosen freely at the input with arbitrary a priori probabilities and information is retrieved from all sequences of photocount events. This additional leeway turns out to improve qualitatively the scaling of mutual information to linear with the signal power when background noise is taken into account.

If the Gaussian model for the background noise is chosen, the result can be compared directly with the ultimate quantum mechanical capacity of a narrowband optical channel. Noticeably, the scaling of optimized generalized on-off keying follows that of the quantum capacity limit in the regime of weak signal power, although a difference in the multiplicative scaling factor indicates that further improvement should be possible by resorting to sub-shot noise quantum receivers.\cite{GuhaHabifJMO2011,ChenHabifNPH2012,BecerraFanNatPhot2013,FerdinandDiMarionpjQI2017}
Another issue emerging from the presented analysis is the very high peak-to-average power ratio required to implement optimal generalized OOK in the deep-space regime, which may have a negative impact on the overall electrical-to-optical conversion efficiency of the transmitter module. However, this issue can be in principle dealt with by implementing recently proposed structured optical receivers which enable efficient optical communication in the photon-starved regime using keying constellation with more uniform instantaneous power distribution.\cite{GuhaPRL2011,RosatiMariPRA2016,BanaszekJachuraICSOS2017}

This paper is organized as follows. In Sec.~\ref{Sec:Operating} we review parameters characterizing a communication link and discuss limits on the communication efficiency. Two models for background  noise are introduced in Sec.~\ref{Sec:NoiseModels}. Optimization of the PPM format in the presence of background noise is carried out in Sec.~\ref{Sec:EfficiencyLimits} with the analysis extended to generalized OOK. An example of a structured optical receiver which may reduce requirements for the peak optical power produced by the transmitter module is discussed in Sec.~\ref{Sec:PowerConsiderations}.
Finally, Sec.~\ref{Sec:Conclusions} concludes the paper.

\section{OPERATING REGIME}
\label{Sec:Operating}

\begin{figure}[t]
\centering
\includegraphics[width=\linewidth]{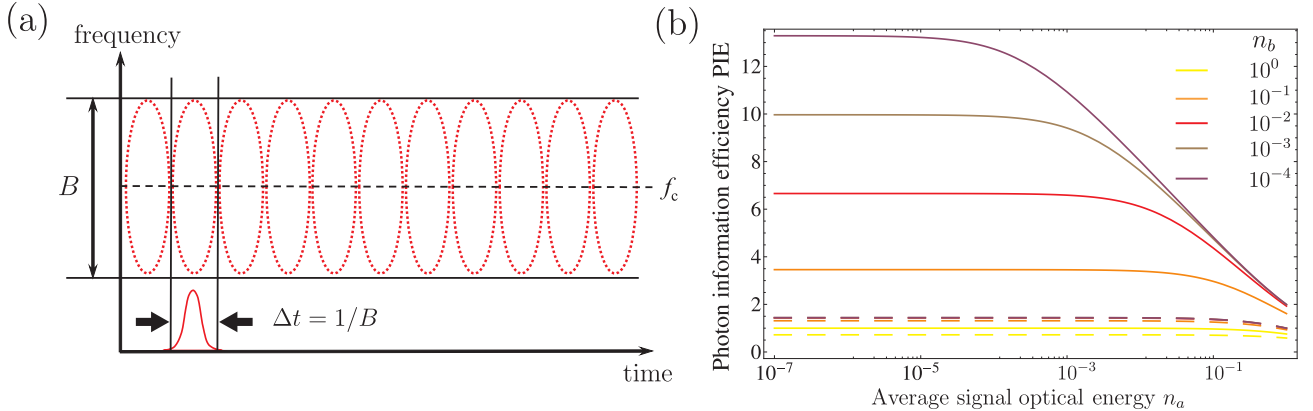}
\caption{(a) Time-frequency diagram of a narrowband communication channel characterized by a carrier frequency $f_c$ and a bandwidth $B$. The channel bandwidth defines the minimum duration $\Delta t = 1/B$ of an individual time bin in which an elementary symbol can be transmitted. (b) Photon information efficiencies for the  Shannon (dashed lines) and the Holevo (solid lines) channel capacities as a function of the average signal optical energy $n_a$ for several values of the noise strength $n_b$ expressed in photons per time bin.\label{Fig:TimeFrequencyDiagram}}
\end{figure}

Parameters characterizing the efficiency of a narrowband optical communication channel can be conveniently discussed using the time-frequency diagram shown in Fig.~\ref{Fig:TimeFrequencyDiagram}(a). Suppose that the transmitter produces a signal with an average optical power $P$ spread over a bandwidth $B$ centered around the carrier frequency $f_c$. The channel bandwidth defines the minimum duration $\Delta t = 1/B$ of an individual time bin in which an elementary symbol can be transmitted. The energy available in a single time bin is therefore equal to $P \Delta t = P/B$. Hence for channel transmission $\eta_{\text{ch}}$ the received optical energy is given by $\eta_{\text{ch}} P / B$. This quantity is usually referred to as the {\em power spectral density} at the channel output. Multiplying it by the efficiency $\eta_{\text{det}}$ of the optical detector yields the detected power spectral density which is a basic figure of merit when calculating the capacity limits. For the purpose of the quantum mechanical analysis of the channel capacity it is convenient to express the average optical energy detected in a single time bin in the units of the energy $hf_c$ of a single photon at the carrier frequency $f_c$, where $h$ is Planck's constant. This yields the dimensionless figure of merit in the form of the average output photon number per one time bin
\begin{equation}
n_a = \eta_{\text{det}} \eta_{\text{ch}}  \frac{ P}{ h f_c B}.
\end{equation}
In conventional optical communication one typically has $n_a \gg 1$. This allows one to encode individual bits of the data stream into a pair of distinct values of the field amplitude and/or phase that can be discriminated at the channel output with a satisfactorily low error probability. For a sufficiently high power, multiple bits can be transmitted in a single time bin using larger constellations of input symbols. In contrast, deep-space optical communication operates in the {\em photon starved regime} when $n_a \ll 1$. This also presents a departure from the typical radio frequency (RF) regime, as presented in Table~\ref{tab:Regimes}.

 \begin{table}
\caption{Examples of operating regimes of radio frequency and optical  communication links for deep-space communication.\cite{Moision2014} Channel transmission $\eta_{\text{ch}}$ is calculated for the distance $R = 1~\mathrm{AU} =  149\cdot 10^9$ m as  $\eta_{\text{ch}} = [ \pi D_{t} D_{r} f_{c}/(4 c R)]^{2} $, where $c=3 \cdot 10^{8}$~m/s is the speed of light in vacuum.}
\label{tab:Regimes}
\begin{center}
\begin{tabular}{|l|l|l|}
\hline
\rule[-1ex]{0pt}{3.5ex}  \textbf{Operating regime} & \textbf{RF} & \textbf{Optical}  \\
\hline
\hline
\rule[-1ex]{0pt}{3.5ex}  Carrier frequency $f_c$ & 32 GHz & $2\cdot 10^5$ GHz   \\
\hline
\rule[-1ex]{0pt}{3.5ex}  Transmit antenna diameter $D_{t}$ & 3 m & 0.22 m  \\
\hline
\rule[-1ex]{0pt}{3.5ex}  Receiver antenna diameter $D_{r}$ & 34 m & 11.8 m  \\
\hline
\rule[-1ex]{0pt}{3.5ex}  Channel transmission $\eta_{\text{ch}}$ & $3.29\cdot 10^{-15}$ & $8.32\cdot 10^{-11}$  \\
\hline
\rule[-1ex]{0pt}{3.5ex}  Detector efficiency $\eta_{\text{det}}$ & 0.1 & 0.025  \\
\hline
\rule[-1ex]{0pt}{3.5ex}  Bandwidth $B$ & 0.5 GHz & 2 GHz  \\
\hline
\rule[-1ex]{0pt}{3.5ex}  Transmit power $P$ & 35 W & 4 W  \\
\hline
\rule[-1ex]{0pt}{3.5ex}  Average output photon number $n_a$ & 1.08 & 0.03  \\
\hline
\rule[-1ex]{0pt}{3.5ex}  Average noise photon number $n_b$ & 66.68 & 0.03  \\
\hline
\rule[-1ex]{0pt}{3.5ex}  Shannon limit transmission rate $B \cdot C_{\text{Shannon}}$  & 11.4 Mbps & 87 Mbps  \\
\hline
\rule[-1ex]{0pt}{3.5ex}  Holevo limit transmission rate $B  \cdot C_{\text{Holevo}} $  & 11.5 Mbps & 273 Mbps \\
\hline
\end{tabular}
\end{center}
\end{table}

The standard Shannon limit on the efficiency of communication using electromagnetic radiation is derived under several assumptions following from the classical wave picture. First, information is encoded in the complex amplitude of the field emitted in consecutive time bins. In the course of transmission this amplitude is affected by additive, phase-insensitive Gaussian noise which adds on average power equivalent to $n_b$ photons per time bin at the detection stage. Finally, the real and imaginary parts of the complex amplitude are simultaneously read out using conventional detection, e.g.\ based on homodyning. Under the constraint of a fixed average detected signal power, equal to $n_a$ photons per time bin, the information-maximizing probability distribution of the
input field amplitude is Gaussian. The effective noise strength at the detection stage is $n_b+1$, where the additional term $1$ comes from the shot noise inherent to homodyne detection. The resulting {\em Shannon capacity} per time bin is given by
\begin{equation}\label{eq:shannon}
C_{\text{Shannon}} = \log_2 \left( 1 + \frac{n_a}{n_b+1} \right)
\end{equation}
which follows directly from the Shannon-Hartley theorem.

The ultimate quantum mechanical limit on the transmission rate takes into account both wave and particle properties of electromagnetic radiation as a carrier of information. This limit is explicitly given by the {\em Holevo capacity} which reads \cite{GiovannettiGarciaPatronNPH2014}
\begin{equation}
C_{\text{Holevo}} = g(n_a+n_b) - g(n_b), \qquad g(x) = (x + 1) \log_2 (x + 1) - x \log_2 x.
\label{Eq:HolevoCapacity}
\end{equation}
Importantly, the Holevo capacity includes optimization over the constellations of symbols (states) used at the channel input and all detection schemes. It is in principle saturable, but may require unconventional measurements beyond direct or homodyne detection.\cite{GuhaHabifJMO2011,ChenHabifNPH2012,BecerraFanNatPhot2013,FerdinandDiMarionpjQI2017}

While for $n_b \gg 1$ the Shannon and the Holevo capacities nearly coincide, as illustrated by the example shown in  Tab.~\ref{tab:Regimes}, there is a substantial difference between these two quantities in the photon-starved regime. When $n_a \ll 1$ the channel capacities tend to zero, which somewhat complicates a discussion of their behavior over a broad parameter range. A convenient solution is to use {\em photon information efficiency} (PIE) defined as $\textrm{PIE}=C/n_a$ which quantifies the maximum information that can be transmitted per time bin per one detected photon. In Fig.~\ref{Fig:TimeFrequencyDiagram}(b) we plot respective PIEs for the Shannon and the Holevo capacities as a function of the signal power $n_a$ for several fixed values of the noise power $n_b$. Conveniently, PIEs tend to constant values  for $n_a \rightarrow 0$.
It is seen that the Shannon capacity is not noticeably affected by the background noise as long as $n_b \ll 1$. This is because in the case of the Shannon capacity the noise floor is effectively defined by the shot noise of homodyne detection. On the other hand, the Holevo capacity takes into account more general detection schemes and the particle nature of electromagnetic radiation, which is essential to achieve efficient communication in the photon-starved regime. Applying the Taylor expansion to Eq.~(\ref{Eq:HolevoCapacity}) yields that in the limit $n_a \rightarrow 0$ the Holevo PIE is  given by
$g'(n_b) = \log_2(1+n_b^{-1})$. The analogous figure for the Shannon capacity is equal to $(\log_2 \textrm{e}) /(1+n_b)$.

As a side note let us note that in telecommunication engineering the term capacity is used to describe the attainable transmission rate per unit time for a given combination of modulation format and detection scheme. Hence it is specified in bits per second. The ultimate quantum mechanical bound on this quantity is given by the product of the Holevo capacity and the channel bandwidth $B$. The capacity per unit bandwidth is referred to as the {\em spectral efficiency} and measured in $\text{bits}/(\text{s} \cdot \text{Hz})$. Hence Eqs.~(\ref{eq:shannon}) and (\ref{Eq:HolevoCapacity}) can be directly used as bounds on the spectral efficiency.

\section{BACKGROUND NOISE}
\label{Sec:NoiseModels}

\begin{figure}[t]
\centering
\includegraphics[width=\linewidth]{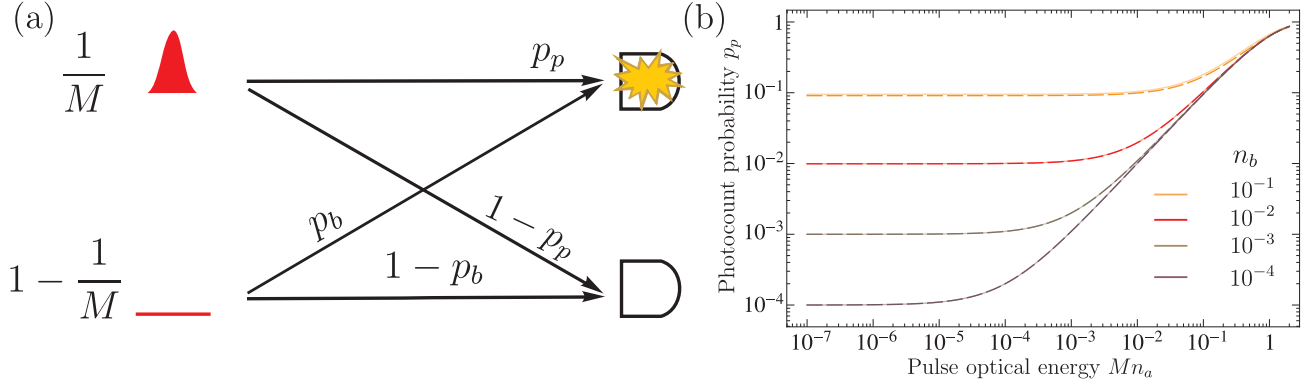}
\caption{(a) Photodetection model. A photocount is generated with conditional probabilities $p_p$ and $p_b$ respectively for the incoming light pulse and an empty time bin. A no-count response is obtained in the remaining cases with respective probabilities $1-p_p$ and $1-p_b$. (b) The probability of generating a photocount for an incoming light pulse as a function of the pulse optical energy for the Poissonian (solid lines) and Gaussian (dashed lines) models of background counts.\label{Fig:probability}}
\end{figure}

The model of the physical layer considered here can be summarized using a diagram presented in Fig.~\ref{Fig:probability}(a). The two elementary single-bin symbols ``on'' (a pulse) and ``off'' (an empty time bin) generate either a photocount or a no-count response on the detector.
For PPM with frames spanning over $M$ time bins the pulse contains the optical energy available for the entire frame, which is equivalent to $M n_a$ photons at the detection stage. In the case of shot-noise limited direct detection the only type of an imperfection is that a pulse will not produce a photocount. According to the standard theory of photodetection,\cite{MandelWolfSemiclPhot} the probability that a photocount is generated is given by $p_p = 1 - e^{-M n_a}$. It is assumed here that the detector does not have the capability to resolve the number of photons that generated a photocount.

In a general noisy scenario one needs to include also a possibility that an empty time bin will result in a photocount.
We will consider two models for noisy signal detection. The first model, referred to as {\em Poissonian} (P), assumes that background counts in a single time bin are described by a Poissonian distribution with a mean $n_b$ and that they are statistically independent from the counts generated by the incoming optical signal. Consequently, the probabilities of generating at least one count by an empty time bin and a light pulse are respectively given by
\begin{equation}
p_b^{(P)} = 1 - e^{-n_b}, \qquad p_p^{(P)} = 1 - e^{-M n_a - n_b}
\label{Eq:PoissonianModel}
\end{equation}
where the product $M n_a$ specifies the mean number of photons carried by the pulse.

The second noise model, called {\em Gaussian} (G), is based on an assumption that the complex amplitude $\alpha$ of the modulated signal prepared in a given time bin experiences additive noise $\alpha \rightarrow \alpha + \beta$, where $\beta$ is a random, complex, phase-invariant Gaussian variable with a mean $\langle\beta\rangle =0$ and the second moment of the absolute value equal to $\langle |\beta|^2 \rangle = n_b$. In this scenario, the count probabilities for an empty time bin and a light pulse read respectively:
\begin{equation}
p_b^{(G)} = \langle 1 - e^{-|\beta|^2} \rangle = \frac{n_b}{n_b+1}, \qquad p_p^{(G)} = \langle 1 - e^{-|\alpha+ \beta|^2} \rangle  = 1 - \frac{e^{-Mn_a/(n_b+1)}}{n_b+1},
\label{Eq:GaussianModel}
\end{equation}
where in the second expression $\alpha$ is the complex amplitude of the signal pulse, $|\alpha|^2 = M n_a$. Expressions for $p_p$ given in Eqs.~(\ref{Eq:PoissonianModel}) and (\ref{Eq:GaussianModel}) have the same expansion up to linear terms in $Mn_a$ and $n_b$. They also coincide for any $Mn_a$ when $n_b=0$. Fig.~\ref{Fig:probability}(b) presents their comparison for the parameter ranges that will be relevant for the further analysis. Although the difference is rather minor, one will be able to see noticeable effects on the transmission rate for the highest noise levels considered here.

The two models specified above follow from different physical pictures of the noise process. The Poissonian model corresponds to a scenario when the radiation seen by the detector comprises multiple modes, each carrying a small fraction of the noise power. Incoherent summation of contributions from these modes results in statistics that can be well approximated by the Poissonian distribution. This model can include also dark counts generated by thermal excitations in the active medium of the photodetector, provided that they are independent from the incoming radiation. It is worth noting that this would not be the case when a non-negligible role is played by the afterpulsing effect. In contrast, the Gaussian model is based on an assumption that in each time bin the detector effectively sees only one mode carrying the modulated signal. Such a regime of operation may be enabled by a selective transfer of the relevant mode to another frequency range,\cite{EcksteinBrechtOpEx2011,BrechtEcksteinNJP2011,ReddyRaymerOpEx2013,BrechtReddyPRX2015,ReddyRaymerOpEx2017}
where it could be detected without the background noise contributed by other orthogonal modes. Transferring individual temporal modes is possible in principle using frequency conversion in non-linear optical media, but this technology still requires maturing before it becomes a viable option. An additional issue would be ensuring synchronization of the incoming signal with the operation of the mode-selective detector. Although mode selectivity should in principle enable substantial reduction of the background noise, in order to facilitate comparison between the two models introduced here we will assume that the background noise parameter $n_b$ is the same in both cases.

\section{EFFICIENCY LIMITS}
\label{Sec:EfficiencyLimits}

The efficiency of a concrete combination of a modulation format and a detection technique can be characterized with the help of  Shannon mutual information which characterizes the strength of correlations between signal preparations and detection outcomes \cite{CoverThomas}. Simple decoding will be considered here, assuming that information is extracted only from events when a photocount occurred in a single time bin within one PPM frame. Detected frames with multiple photocounts are interpreted as erasures. Optimization is carried out over the size $M$ of the PPM frame treated as a continuous real parameter, which is admissible as long as the optimal $M^\ast \gg 1 $.

\begin{figure}[t]
\centering
\includegraphics[width=\linewidth]{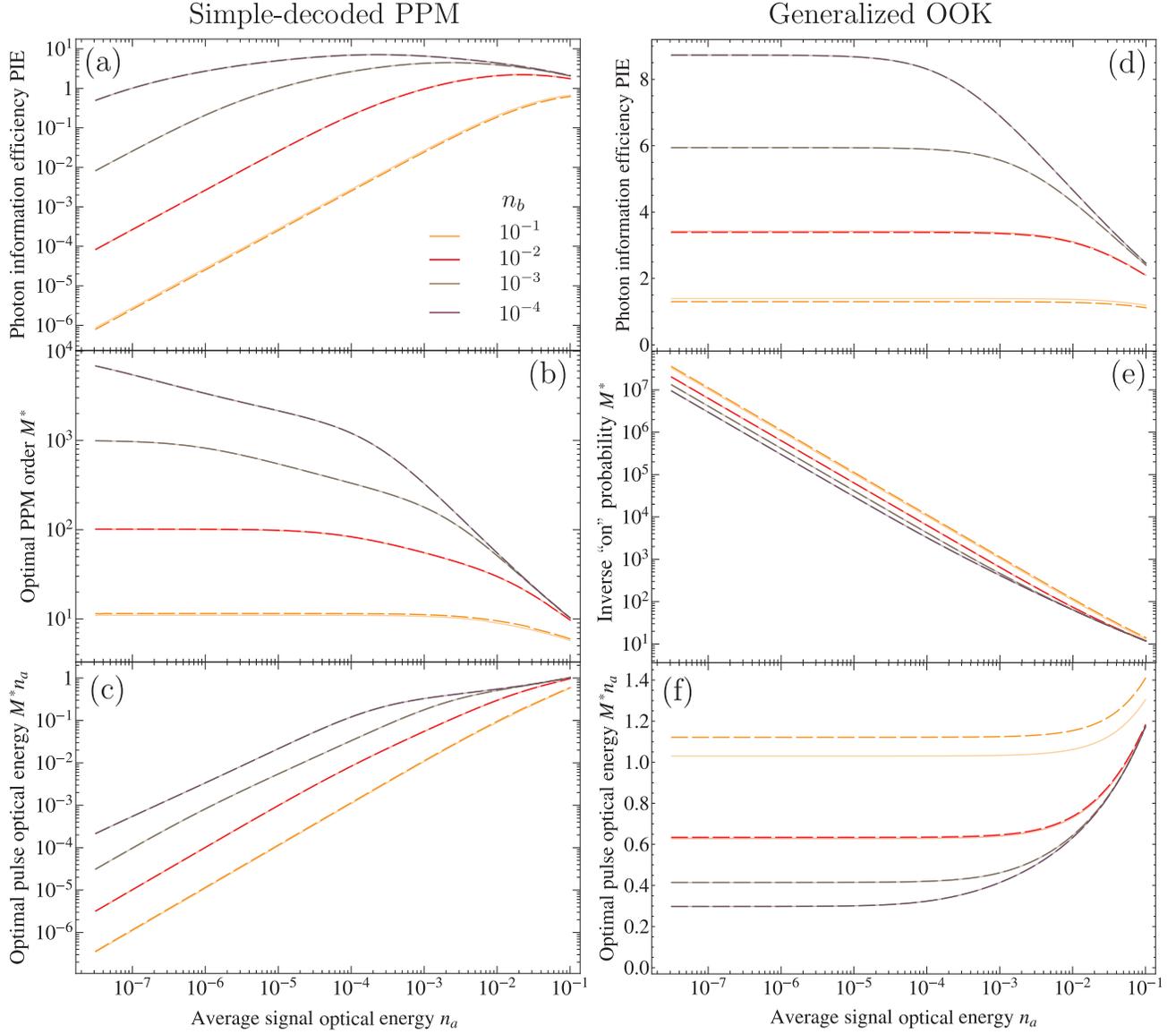}
\caption{Optimized communication using a binary constellation of ``on'' and ``off'' symbols for simple decoded PPM (a-c) and  generalized OOK (d-f) assuming the background noise power equivalent to $n_b = 10^{-1}, 10^{-2}, 10^{-3}, 10^{-4}$ detected photons per time bin. The PIE attainable in both scenarios is shown in (a,d). In the case of PPM optimization is carried out over the frame length $M$ with results shown in (b), which for generalized OOK corresponds to the inverse of the probability of using the ``on'' symbol depicted in (e). The optimal detected optical energy of a pulse is plotted in (c) and (f). Results are shown for the Poissonian (solid lines) and the Gaussian (dashed lines) models of the background noise.}
\label{Fig:OptimizedPPMandOOK}
\end{figure}
%Mn_a= PPM goes to 0 proportionally to n_a, OOK - saturates at constant value with decreasing n_a; PIE=PPM goes to 0 proportionally to n_a, OOK - saturates at constant value with decreasing n_a.

Fig.~\ref{Fig:OptimizedPPMandOOK}(a) depicts the optimal PIE for simple-decoded PPM with direct detection optimized over the frame length. It is known that in the noiseless scenario the PIE grows unlimited with the diminishing signal power provided that the PPM frames can be arbitrarily long.\cite{KochmanWangTIT2014,JarzynaKuszajOPEX2015} In the presence of noise, Fig.~\ref{Fig:OptimizedPPMandOOK}(a) shows that for a given noise power $n_b$ such growth can be observed for signal powers $n_a \gtrsim n_b$. In the opposite regime, when $n_a \lesssim n_b$, the PIE decreases with the diminishing signal power. In the limit $n_a \rightarrow 0$ it exhibits approximately linear scaling with $n_a$. As seen in Fig.~\ref{Fig:OptimizedPPMandOOK}(b), in this limit the optimal PPM frame length $M^\ast$ saturates at a constant value dependent on the background noise power. Consequently, the detected optical energy of the pulse, given by $M^\ast n_a$ and shown in Fig.~\ref{Fig:OptimizedPPMandOOK}(c), tends to zero when $n_a \rightarrow 0$. As a result, received pulses introduce only a minor modification in the statistics of the generated photocounts which becomes dominated by the background noise. The above conclusions hold irrespectively of the Poissonian or the Gaussian model for the background noise employed in the calculations.

A binary alphabet consisting of a pulse and an empty time bin can be used to implement generalized OOK, where the two input symbols can be used with different a priori probabilities. Denoting these probabilities respectively by $1/M$ and $1-1/M$ allows one to establish a direct correspondence with the PPM format in which a pulse is sent exactly once in each frame of $M$ otherwise empty time bins. From the information-theoretic viewpoint, generalized OOK with direct detection is described by a binary asymmetric channel with conditional probabilities depicted in Fig.~\ref{Fig:probability}(a). In order to investigate the full potential of this scheme, we will carry out optimization of generalized OOK over $M$ without any constraint on the maximum optical energy of the pulse. According to Fig.~\ref{Fig:OptimizedPPMandOOK}(d), in this scenario the PIE exhibits a qualitatively different behavior compared to the PPM case, as it tends to a non-zero constant value when $n_a \rightarrow 0$. Fig.~\ref{Fig:OptimizedPPMandOOK}(e) shows that in this limit the pulses are optimally sent less and less frequently. This is related to the fact that the optimal detected optical energy of the pulse stays at the level of the order of one photon even for an arbitrarily low average signal power as seen in Fig.~\ref{Fig:OptimizedPPMandOOK}(f). Consequently, a bin containing a pulse generates a photocount with a much higher probability than an empty one. This observation provides intuition behind the qualitatively different behavior of PIE for generalized OOK compared to the PPM scheme considered before.

\begin{figure}[t]
\centering
\includegraphics[width=\linewidth]{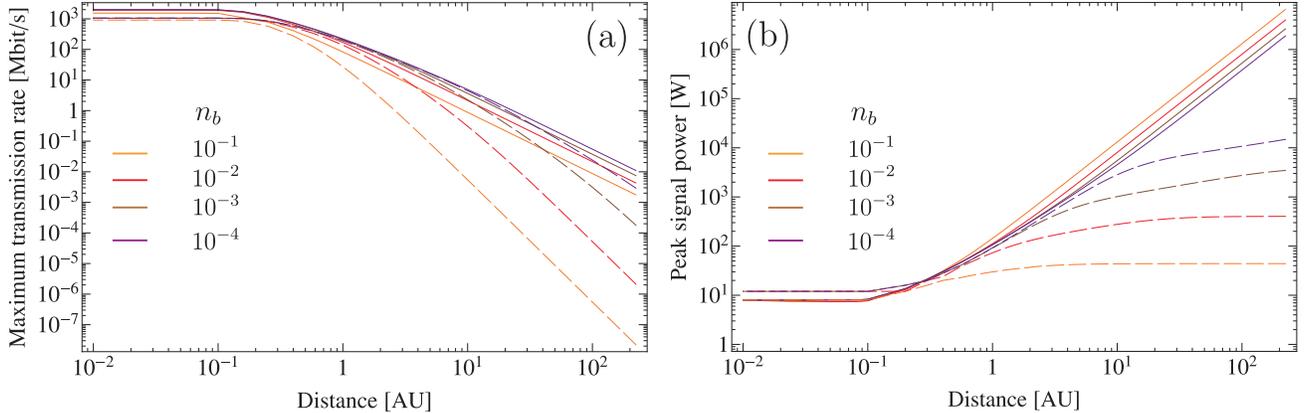}
\caption{The maximum attainable transmission rate (a) and the required transmitter peak power (b) as a function of distance expressed in astronomical units, $1~\textrm{AU}=149\cdot10^6~\textrm{km}$, for the parameters of an optical link specified in Tab.~\ref{tab:Regimes}. Results for simple-decoded PPM (dashed lines) are compared with generalized OOK (solid lines). The four values of the background noise parameter $n_{b} = 10^{-1}, 10^{-2}, 10^{-3}, 10^{-4}$ correspond to the detected noise power of $26.5~\mathrm{pW}$, $2.65~\mathrm{pW}$, $0.265~\mathrm{pW}$, and $0.0265~\mathrm{pW}$ respectively.}
\label{Fig:TransmissionRate}
\end{figure}

The advantage offered by generalized OOK without the constraints of simple-decoded PPM has a dramatic effect on the transmission rates that could be attained in deep-space optical communication. This is illustrated with Fig.~\ref{Fig:TransmissionRate}(a) which depicts the maximum transmission rate as a function of the distance assuming numerical parameters of an optical communication link specified in Tab.~\ref{tab:Regimes}. For large distances generalized OOK can offer rates scaling $R^{-2}$ with the distance $R$ in contrast to conventional PPM which exhibits inferior $R^{-4}$ dependence. This is a consequence of different behaviour of the PIE, tending to a constant for $n_a \rightarrow 0$ in the former case, and proportional to $n_a$ in the latter case.
As a simple illustration, for the noise power equal to $2.56~\textrm{pW}$, corresponding to $n_b=10^{-2}$, generalized OOK can deliver in principle almost a tenfold gain in the transmission rate compared to PPM at the distance $R=10~\textrm{AU}$.
Of course, Shannon mutual information is only the upper bound on the transmission rate. Rates that can be reached in practice depend essentially on forward error correction, which may be much more challenging to implement efficiently in the case of generalized OOK.\cite{HemmatiBiswasProcIEEE2011}

In the case of the Gaussian model for the background noise, the asymptotic values of the PIE for $n_a \rightarrow 0$ seen in Fig.~\ref{Fig:OptimizedPPMandOOK}(d) can be directly compared with the the PIEs resulting from the Holevo capacity which can be read out from Fig.~\ref{Fig:TimeFrequencyDiagram}(b). It is seen that the latter figures are consistently higher. It has been proven using a quantum information theoretic approach\cite{Jarzyna2017,DingPavlichinXXX2017} that under the constraint of a fixed average signal power the constellation composed of an empty time bin and a coherent pulse saturates the Holevo capacity in the limit $n_a \rightarrow 0$. As the Holevo capacity incorporates optimization over all physically permissible detection strategies, the gap between the asymptotic values seen in Fig.~\ref{Fig:TimeFrequencyDiagram}(b) and Fig.~\ref{Fig:OptimizedPPMandOOK}(d) implies that the efficiency of generalized OOK can be further enhanced using more sophisticated receivers.\cite{GuhaHabifJMO2011,ChenHabifNPH2012,BecerraFanNatPhot2013,FerdinandDiMarionpjQI2017}
It is worth noting that for direct detection generalized OOK  achieves the highest possible information rate in the asymptotic limit of the diminishing average signal power.\cite{Verdu1990}

\begin{figure}
	\begin{center}
		\includegraphics[width=0.8\textwidth]{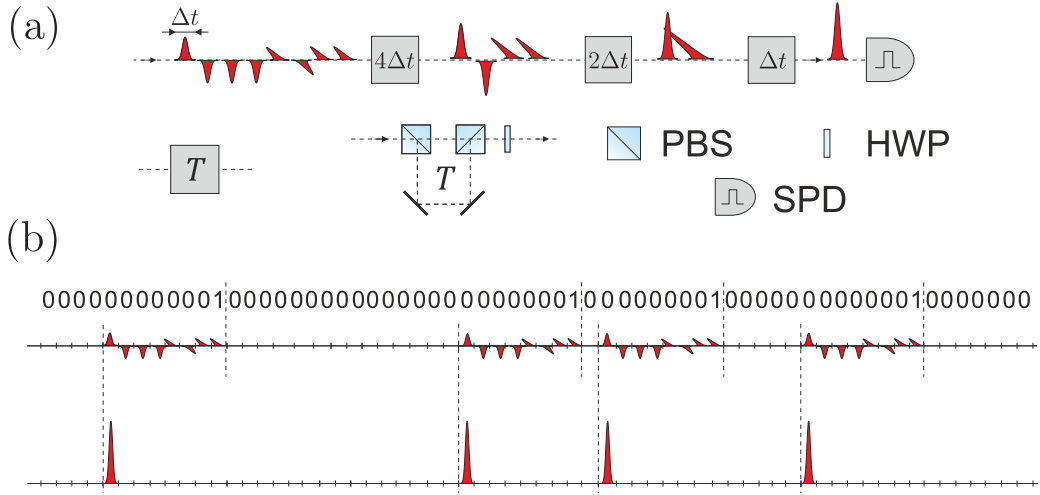}
\vspace{3mm}
		\caption{(a) An exemplary structured optical receiver which concentrates optical energy carried by a phase-polarization pattern into a single time bin of duration $\Delta t$. The choice between the horizontal and the vertical polarizations as well as $0$ and $\pi$ phases is represented by the orientation of pulses in individual bins. The receiver is composed of a sequence of modules. Each module contains a polarization-selective delay line constructed using two polarizing beam splitters PBS to overlap the earlier half of the pattern  with the later half. For a suitable choice of the relative phase, the resulting superpositions are polarized at $\pm 45^\circ$, which are brought to the rectilinear basis using a half-wave plate HWP. After the last module the entire optical energy is concentrated in one time bin and detected using a single photon detector SPD. (b) The temporal location of the produced pulses is synchronized with the arrival of phase-polarization patters. In order to ensure rotational invariance between the transmitter and the receiver the patterns should be prepared using the circular rather than the rectilinear polarization basis and converted into the latter one at the receiver entrance using a quarter-wave plate.} \label{Fig:LoopyPolarization}
	\end{center}
\end{figure}

\section{POWER CONSIDERATIONS}
\label{Sec:PowerConsiderations}

The advantage of generalized OOK for deep-space optical communication discussed in the preceding section relies on the hardware ability to send infrequently light pulses with high optical energy. Fig.~\ref{Fig:TransmissionRate}(b) illustrates the requirements for the transmitter peak power as a function of distance assuming a simple rectangular pulse shape completely filling the time bin. Standard solutions for high-order PPM laser transmitters are based on a master oscillator power amplifier (MOPA) architecture or diode pumped solid-state lasers employing Q-switching technique. Such systems are able to provide peak powers respectively of several and tens of kilowatts while maintaining the optical pulse duration below $1~\mathrm{ns}$. However, in both cases electrical power consumption becomes problematic. Additionally Q-switched laser systems suffer from a limited repetition rate.\cite{Hemmati,Caplan} Owing to spacecraft power limitations and heat dissipation issues, the peak power achievable by the transmitter laser system is usually expected to remain in the range of few kilowatts. In the numerical example shown in Fig.~\ref{Fig:TransmissionRate}(b) this would imply suboptimal performance for ranges exceeding several astronomical units.

The requirement for high peak power stems from the need to confine the signal-generated photocounts to short time intervals. One can envisage a technique to send the signal in a temporally extended form that would undergo optical compression before  photodetection. The advantage of this approach may be more efficient generation of the optical signal in the transmitter assembly utilizing e.g.\ the MOPA architecture. One possibility to implement this approach would be to generate chirped pulses and subject them to compression in a dispersive medium in the first stage of the receiver setup. An alternative approach proposed recently\cite{GuhaPRL2011,RosatiMariPRA2016,BanaszekJachuraICSOS2017} is based on sequences of phase- and (optionally) polarization-modulated pulses whose optical energy can be concentrated in a single time bin using an interferometric setup. This idea is illustrated with an exemplary design for a structured optical receiver shown in Fig.~\ref{Fig:LoopyPolarization}(a). The transmitter emits phase-polarization pulse patterns spanning over multiple time bins whose number is an integer power of $2$. The receiver consists of a series of modules. Each module introduces a relative delay between the two input polarizations equal to half of the duration of the pattern entering that module and rotates the output polarization by $45^\circ$ using a half-wave plate. Phases and polarizations of individual pulses in the input pattern are chosen such that after each module the length of the pattern is reduced by a factor of $2$.
After the last module the entire optical energy of the pattern is concentrated in a single time bin whose temporal location is synchronized with the arrival time of the pattern as shown schematically in Fig.~\ref{Fig:LoopyPolarization}(b). Hence direct detection can be used to read out timing information with the resolution of an individual time bin. Note that the use of phase-polarization patters requires a minimum separation between consecutive pulses at least equal to the pattern length. This however should have a minor effect on the efficiency.

The reduction of the required peak-to-average power ratio for the signal emitted by the transmitter may improve its overall electrical-to-optical conversion efficiency. The disadvantage of this solution is a more complicated construction of the receiver, which in the example discussed above requires maintaining interferometric stability between multiple optical paths. The interferometric setup would also need to tolerate spatial distortions of the signal introduced e.g.\ by atmospheric turbulence.\cite{JinAgneXXX2015} It remains to be investigated whether inevitable imperfections of structured receivers would realistically cancel the advantages of improved trasmitter power management. Results obtained so far indicate robustness to certain types of imperfections.\cite{JarzynaLipinskaOPEX2016} Alternative designs for structured optical receivers may become viable with the development of fast, low-loss optical switches\cite{BanaszekJachuraICSOS2017} as well as quantum memories.\cite{KlimekJachuraJMO2016}

\section{CONCLUSIONS}
\label{Sec:Conclusions}

The overall efficiency of an optical communication link depends on a number of hardware and software factors. The purpose of the present paper was to discuss the potential of the physical layer for deep-space optical communication in the presence of background noise. The theoretical reference in the analysis was the quantum mechanical Holevo channel capacity which takes into account both wave and particle aspects of electromagnetic radation as the carrier of information. With the diminishing average signal power generalized OOK was found to deliver in principle much better performance than the standard simple-decoded PPM, provided that the two OOK symbols are used with highly imbalanced a priori probabilities. Optimal implementation of generalized OOK requires concentration of the optical energy in very few time bins, which leads to demanding requirements for the peak-to-average power ratio of the transmitter laser source. This problem can be in principle alleviated by the use of structured receivers which exploit optical interference to concentrate temporally optical power. More complicated architectures of such receivers may not be a critical issue for downlink transmission as the size, weight, and power consumption are secondary considerations in the case of ground terminals. Exploiting the full potential of the physical layer may also require further advancements in the area of error correction to approach with practical codes Shannon mutual information for transmission schemes more general than PPM.

\acknowledgments

We acknowledge insightful discussions with F. E. Becerra, S. Guha, C. Heese, Ch. Marquardt, and J. Nunn.
This work is part of the project ``Quantum Optical Communication Systems'' carried out within the TEAM
programme of the Foundation for Polish Science co-financed by the European Union under the European
Regional Development Fund.

\bibliography{deepspace} % bibliography data in report.bib

\bibliographystyle{spiebib} % makes bibtex use spiebib.bst

%%\begin{thebibliography}{99}

\end{document}